\documentclass[aps,prl,reprint,amssymb,10pt]{revtex4-2}

\usepackage{amsmath,amssymb,color,braket,enumitem}
\usepackage{bm}
\usepackage{hyperref}
\usepackage{graphicx}
\usepackage[dvipsnames]{xcolor}

\DeclareMathOperator{\Tr}{Tr}

\begin{document}

\title{  
Bad Metal Behavior and Lifshitz Transition of a
Nagaoka Ferromagnet}

\author{Jonas Arnold}
\affiliation{Institut f\"{u}r Theoretische Physik, Universit\"{a}t Frankfurt,  Max-von-Laue Stra{\ss}e 1, 60438 Frankfurt, Germany}

\author{Peter Kopietz}
\affiliation{Institut f\"{u}r Theoretische Physik, Universit\"{a}t Frankfurt,  Max-von-Laue Stra{\ss}e 1, 60438 Frankfurt, Germany}

\author{Andreas R\"{u}ckriegel}
\affiliation{Institut f\"{u}r Theoretische Physik, Universit\"{a}t Frankfurt,  Max-von-Laue Stra{\ss}e 1, 60438 Frankfurt, Germany}

\date{August 28, 2026}

\begin{abstract}
%
%

Using an extension of the fermionic functional renormalization group
for systems where strong correlations give rise to projected Hilbert spaces 
we
calculate the phase diagram and the electronic spectral function 
of the Hubbard model at infinite on-site repulsion.
For a square lattice with nearest-neighbor hopping
we find that the
ground state evolves from a 
paramagnetic Fermi liquid at low densities 
via a state with antiferromagnetic stripe order at intermediate densities 
to an extended Nagaoka ferromagnet at high densities. 
The single-particle spectral function of the
Nagaoka ferromagnet exhibits a flat but rather broad band characteristic for an incoherent non-Fermi liquid. We identify two distinct ferromagnetic regimes
separated by a Lifshitz transition.

%
%
%
%
\end{abstract}

\maketitle

\textit{Introduction---}One of the few exact results for the Hubbard model in dimensions greater than one is the Nagaoka
theorem \cite{Nagaoka1965, Nagaoka1966, Kollar1996, Tasaki1998},  which states that for
nearest-neighbor hopping on a bipartite lattice with periodic boundary conditions
and infinite on-site repulsion the ground state of the Hubbard model doped with a single hole away from half filling is ferromagnetic. This so-called kinetic or Nagaoka ferromagnetism is stabilized by path interference: only in a ferromagnetic state can a single hole move freely, without disturbing the background spin texture and creating defects. This enables constructive interference of different hopping paths on bipartite lattices,  thereby lowering the kinetic energy. 
The Nagaoka theorem is even more surprising when considering an equally exact theorem by Lieb \cite{Lieb1989},
which states that the ground state of the Hubbard model on a bipartite lattice at half filling is antiferromagnetic for arbitrary on-site repulsion $ U $.
Consequently,
the stability of Nagaoka ferromagnetism for more than one hole, 
and especially for a finite doping concentration in the thermodynamic limit, 
has been subject to intense and ongoing research \cite{Roth1967, Brinkman1970,  Aruiac1990, Elser1990, Kotrla1990, Izyumov1990, Shastry1990, vonderLinden1991, Putikka1992, Hanisch1993, Chiappe1993, Zotos1993, Wurth1995, Kollar1996, Kuzmin1997, Obermeier1997, Tasaki1998, Becca2001, White2001, Coleman2002, Zitzler2002, Park2008, Baroni2011, Liu2012, Kochetov2017, Blesio2019, Morera2023, Samajdar2024a, Newby2025, Sharma2025}.
Recent experimental progress realizing the infinite-$U$  Hubbard model on optical lattices only reinvigorates the interest in this question \cite{Cheuk2016, Dehollain2019, Bohrdt2021, Spar2021, Lebrat2024, Prichard2024, Kendrick2025}.
However,
the theoretical analysis of the infinite-$U$ Hubbard model  and its Nagaoka ferromagnetism has proven challenging.
The main difficulty is that the correct description of the dynamics of holes requires very large system sizes \cite{White2001, Maska2012}.
This is already obvious from the Nagaoka theorem itself,
which shows that the presence of a single hole completely changes the spin configuration of the entire system,
at arbitrary distances.
The resulting dependence on lattice size and boundary conditions \cite{Aruiac1990, White2001, Kochetov2017, Zotos1993} severely limits the applicability and predictive power of numerical methods such as  exact diagonalization, density-matrix renormalization group, and Monte Carlo simulations.

Since the Nagaoka state is a strongly correlated  itinerant ferromagnet, the electronic single-particle spectral function must be non-trivial. Surprisingly, this important  
quantity has been largely ignored in the literature, perhaps due to the lack of controlled methods for calculating spectral functions of electronic lattice models in the extremely correlated  regime \cite{Shastry2010,Shastry2011,Shastry2013} where the on-site repulsion is large compared to all other energy scales. Physical properties at all relevant energy scales 
are then determined by states in  a projected Hilbert space consisting of all states without
doubly occupied lattice sites. 
Unfortunatly, the Hilbert space projection cannot be implemented perturbatively, so that
so far it has not been possible to study Nagaoka ferromagentism using
perturbative weak-coupling expansions or more sophisticated resummations of 
perturbation theory such as various implementations of the
functional renormalization group (FRG) for interacting fermions~\cite{Salmhofer2001, Honerkamp2001, Kopietz2001, Kopietz2010, Metzner2012, Dupuis2021,Halboth2000, Halboth2000b, Salmhofer2004, Ossadnik2008, Husemann2009, Husemann2012, Taranto2014, Lichtenstein2017, Vilardi2017, Honerkamp2018, Vilardi2019, Ehrlich2020, Hille2020, Honerkamp2022}.

\begin{figure}%
\includegraphics[width=\linewidth]{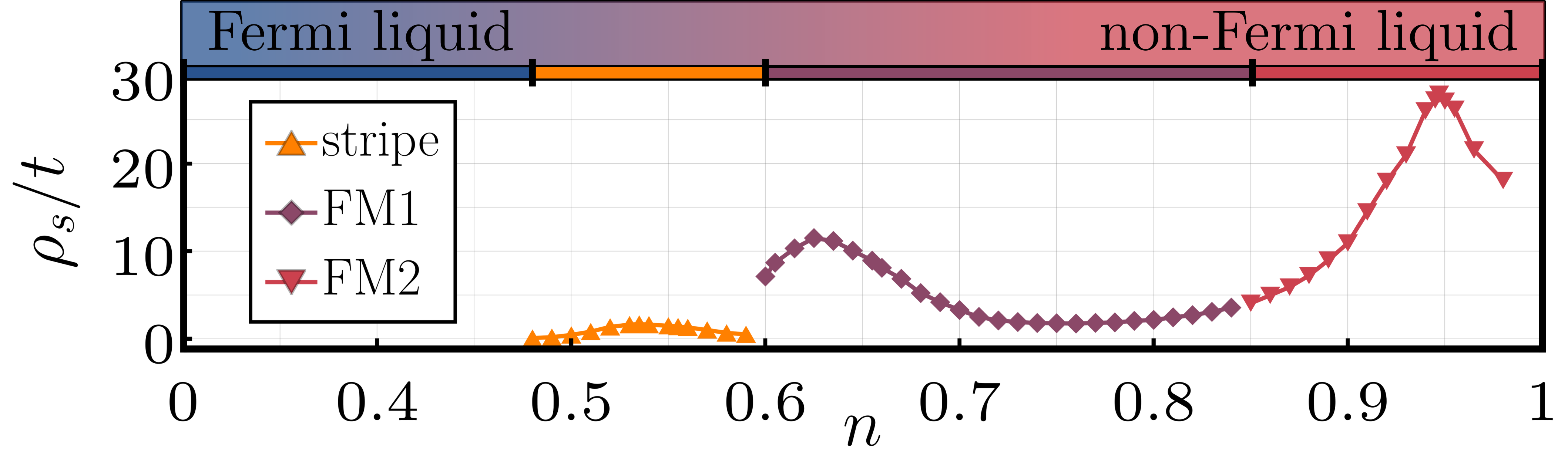}
\caption{Ground state phase diagram of the $ t $ model (top).
With increasing density $ n $,
the system transitions from a paramagnet (blue) to stripe magnet (orange) to ferromagnet (purple and red).
This is accompanied by a  crossover from a Fermi liquid at small $ n $ to an incoherent non-Fermi liquid at large $ n $.
The bottom panel shows the spin stiffness $ \rho_s $ of the magnetic instabilities,
showing two distinct ferromagnetic peaks that we associate with two different phases (FM1 and FM2) separated by a Lifshitz transition. 
}
\label{fig:phases}
\end{figure}

In this Letter we show that this limitation of the FRG can be transcended
by  our recently developed \cite{Rueckriegel2023} formulation of the FRG in terms of Hubbard X-operators \cite{Fulde1995,Izyumov1988,Fazekas1999,Ovchinnikov2004}
dubbed X-FRG.  With this method,  the strong correlations implied by the
Hilbert space projection are taken into account exactly by imposing a non-trivial initial condition on the FRG flow. This enables us to calculate not only the
phase diagram of the $U = \infty$ Hubbard model as a function of doping, but also the entire single-particle spectral function.  In the high-doping regime where the ground state is ferromagnetic we find that the spectral function exhibits 
an incoherent flat band which can be considered as a signature  of strong correlations. 
Our final result for the ground state phase diagram of the $U= \infty$ Hubbard model
is shown in Fig.~\ref{fig:phases}.
In agreement with most earlier studies \cite{Shastry1990, Izyumov1990, vonderLinden1991, Hanisch1993, Wurth1995, Obermeier1997, Becca2001, Park2008, Baroni2011, Liu2012, Blesio2019, Newby2025}, at high densities we find 
an extended Nagaoka ferromagnet phase while at low densities the system is a paramagnetic Fermi liquid. 
Additionally, 
we discover that these phases are separated by a finite interval of intermediate densities where the ground state exhibits antiferromagnetic stripe order.
Moreover, our  independent calculation of the single-particle spectral function reveals that the high-density Nagaoka phase can be subdivided  into two distinct regimes
separated by a Lifshitz transition where the topology of the Fermi surface changes discontinuously.

As pointed out by Anderson \cite{Anderson2008, Anderson2009, Casey2011},
to understand the effect of the Hilbert space projection on the single-particle excitations  
it is sufficient to consider  the limit of infinite on-site repulsion 
$U \rightarrow \infty$ in the Hubbard model. The Hamiltonian is then given by the  so-called $t$ model, i.e., the projected kinetic energy
\begin{equation}\label{eq:t_model}
	\mathcal{H} 
	=
	\sum_{ i j } \sum_\sigma
	t_{ i j }
	h_{ i \sigma }^\dagger 
	h_{ j \sigma } 
	- \mu \sum_i  \sum_\sigma
	n_{ i \sigma }
	,
\end{equation}
where    $\mu$ is the chemical potential and we consider nearest neighbor hopping on a square lattice,
i.e. $ t_{ i j } = t $  for all pairs  $ i , j $ of
nearest neighbors and $ t_{ i j } = 0 $ otherwise. 
The holon operators
$h_{i \sigma}$ and $h^{\dagger}_{ i \sigma}$
act on the projected Hilbert space consisting of fermionic Fock states
without doubly occupied lattice sites. In this space 
$h_{i \sigma}$ and $h^{\dagger}_{ i \sigma}$
annihilate and create a spin-$\sigma$ electron at site $i$, while
$ n_{ i \sigma } = 
h_{ i \sigma }^\dagger 
h_{ i \sigma }  $
counts the number of spin-$\sigma $ electrons in a grand 
canonical ensemble with chemical potential $\mu$.
Despite its deceptively simple quadratic form,
the Hamiltonian \eqref{eq:t_model} is highly non-trivial because 
$ h_{ i \sigma } $
creates a hole in a restricted Hilbert space without double 
occupancy.
As a consequence, $h_{i \sigma}$ and $h^{\dagger}_{ i \sigma}$ are  
\textit{not} canonical fermion operators but
obey 
the non-canonical anticommutation relations 
\begin{equation}
h_{ i \sigma } 
h_{ j \sigma' }^\dagger  
+
h_{ j \sigma' }^\dagger  h_{ i \sigma } 
=
\delta_{ i j } [
\delta_{ \sigma \sigma' }
\left(
1 - n_{ i \bar{ \sigma } }
\right)
+ 
\delta_{ \sigma \bar{ \sigma }' }
h_{ i \bar{ \sigma } }^\dagger 
h_{ i \sigma } ]  ,
\label{eq:holon_algebra}
\end{equation}
where $ \bar{\sigma} = - \sigma $.
The relation \eqref{eq:holon_algebra} 
encodes the restricted local Hilbert space,
and in particular entails the constraint
$ n_{ i \uparrow }
+ n_{ i \downarrow }
= 0 $ or $ 1 $.

\textit{X-FRG for the $ t $ model---}The X-FRG \cite{Rueckriegel2023} extends  the spin FRG proposed in Ref.~[\onlinecite{Krieg2019}] 
(for recent applications see Refs.~[\onlinecite{Rueckriegel2024a, Rueckriegel2024b}])
to more general models for strongly correlated electrons.
The basic idea is to use a non-trivial local deformation of the original model as initial condition for the FRG 
flow \cite{Machado2010, Rancon2011a, Rancon2011b}.
The flow equations provide an unbiased and non-perturbative resummation of non-local correlations which allows us to apply the full power of the established FRG machinery \cite{Berges2002, Pawlowski2007, Kopietz2010, Metzner2012, Dupuis2021}
to lattice models defined on restricted Hilbert spaces.
The initial condition of the X-FRG flow is the exactly solvable $ t = 0 $ limit of isolated Hubbard atoms.
Hopping-induced correlation effects are incorporated by a non-perturbative resummation of the series in $ t / T $,
where $ T $ is the temperature.
This is achieved by replacing 
$ t \to t_\Lambda = \Lambda t $ and following the evolution of the effective holon action from $ \Lambda = 0 $ to $ 1 $,
which is equivalent to lowering the temperature from infinity to $ T = 1 / \beta $ \cite{Rueckriegel2024a,Arnold2025}.
The starting point is the generating functional of imaginary-time-ordered connected holon correlation functions,
\begin{align}
&
e^{ \mathcal{G}_\Lambda [ \bar{j} , j ] }
= 
\Tr \left\{
e^{ 
\beta \mu_0 \sum_{ i \sigma } n_{ i \sigma }
}
\mathcal{T}
e^{
- \int_0^\beta \textrm{d} \tau
\sum_{ i j \sigma } 
t_{ \Lambda , i j } 
h_{ i \sigma }^\dagger ( \tau )
h_{ j \sigma } ( \tau )
}
\right.
\nonumber\\
& \hspace{3mm} 
\left. \times
e^{
\int_0^\beta \textrm{d} \tau
\sum_{ i \vphantom{j} \sigma } 
\left[
\delta \mu_\Lambda
h_{ i \sigma }^\dagger ( \tau )
h_{ i \sigma } ( \tau )
+
\bar{j}_{ i \sigma } ( \tau )
h_{ i \sigma } ( \tau )
+
h_{ i \sigma }^\dagger ( \tau )
j_{ i \sigma } ( \tau )
\right]
}
\right\} ,
\label{eq:generating_functional_G}
\end{align}
where
$ h_{ i \sigma } ( \tau ) =
e^{ \mu_0 \tau } h_{ i \sigma } $,
$ \mathcal{T} $
denotes imaginary-time ordering,
and
$ j_{ i \sigma } ( \tau ) $ and
$ \bar{j}_{ i \sigma } ( \tau ) $
are Grassmann source fields.
Note that we also include a counter-term 
$ \delta \mu_\Lambda $ for the chemical potential.
Hence, 
$ \mu = \mu_0 + \delta \mu_\Lambda $,
with the Hubbard atom initial value
$ \mu_0 = T \ln [ n / ( 2 - 2 n ) ] $
for a given density $ n $.
The counter-term allows us to keep the density $ n $ fixed as $ \Lambda $ varies;
this is advantageous because $ n $ controls the high-frequency behavior of the holon propagator $G_{\Lambda} ( K )$.
In Ref.~[\onlinecite{Rueckriegel2023}],
we have shown that
$ \mathcal{G}_\Lambda [ \bar{j} , j ] $
satisfies a standard fermionic FRG flow equation, from which we can obtain the
usual fermionic Wetterich equation via Legendre transformation  \cite{Salmhofer2001, Honerkamp2001, Kopietz2001, Kopietz2010, Metzner2012, Dupuis2021}. 
Hence,
we can apply the established machinery of the fermionic FRG 
\cite{Salmhofer2001, Honerkamp2001, Kopietz2001, Kopietz2010, Metzner2012, Dupuis2021,Halboth2000, Halboth2000b, Salmhofer2004, Ossadnik2008, Husemann2009, Husemann2012, Taranto2014, Lichtenstein2017, Vilardi2017, Honerkamp2018, Vilardi2019, Ehrlich2020, Hille2020, Honerkamp2022} to the $ t $ model \eqref{eq:t_model}.
To that end,
we parametrize the holon propagator as \cite{Rueckriegel2023,Arnold2025}
\begin{equation}
\label{eq:holon_propagator}
G_\Lambda ( K ) 
=
\frac{ Z }{ 
i \omega + \mu_0 
- Z \left( 
t_{ \Lambda , \bm{k} }
- 2 \delta \mu_\Lambda / 3 
\right)
- Z \Sigma_\Lambda ( K )
} ,
\end{equation}
where $ K = ( \bm{k} , \omega ) $ collects crystal momentum $ \bm{k} $ and fermionic Matsubara frequency $ \omega $,
the quasi-particle residue of the Hubbard atom is $ Z = 1 - n / 2  $,
and the holon self-energy
$ \Sigma_\Lambda ( K ) $ encodes effects of $ t_\Lambda $ beyond tree level.
%
%
%
Its flow is governed by the 
two-body interaction vertex
$ U_\Lambda ( K_1' , K_2' ; K_2 , K_1 ) $,
which is itself determined by the particle-particle (pp),
forward-scattering (fs), and 
exchange (ex) ladder diagrams and the three-body interaction vertex,
which we neglect. For the explicit form of the relevant flow equation  we refer to the companion paper \cite{Arnold2025}.
Following the methodology of fermionic FRG,
we apply a channel decomposition \cite{Husemann2009, Metzner2012, Vilardi2017} to the two-body interaction,
\begin{align}
&
U_\Lambda ( K_1' , K_2' ; K_2 , K_1 )
\nonumber\\
= {} &  
  \mathcal{V} ( \omega_2 , \omega_1 )
- \mathcal{S}_\Lambda ( Q_\textrm{pp} ; \omega_2' , \omega_1 )
+ \mathcal{M}_\Lambda ( Q_\textrm{ex} ; \omega_2 , \omega_1 )
\nonumber\\
& 
+ \frac{ 1 }{ 2 } \Bigl[
\mathcal{M}_\Lambda ( Q_\textrm{fs} ; \omega_1 , \omega_2 ) -
\mathcal{C}_\Lambda ( Q_\textrm{fs} ; \omega_2 , \omega_1 )
\Bigr] .
\label{eq:U_channels}
\end{align}
Here,
$ \mathcal{S}_\Lambda $,
$ \mathcal{M}_\Lambda $, and
$ \mathcal{C}_\Lambda $
are the superconducting,
magnetic, 
and charge channels,
respectively,
$ Q_{ x } = ( \bm{q}_x , \Omega_x ) $,
$ x \in \{ \textrm{pp} , \textrm{fs} , \textrm{ex} \} $
denote the relevant (bosonic) momentum-frequency transfers,
and
$ \mathcal{V} $ 
is a $ Q_x $-independent part of the Hubbard atom vertex.
Due to the holon algebra \eqref{eq:holon_algebra},
the channels have non-trivial initial conditions,
\begin{subequations} \label{eq:U0}
\begin{align}
Z^2\mathcal{V} ( \omega_2 , \omega_1 ) 
= {} &
- G_0^{ - 1 } ( \omega_2 )
- G_0^{ - 1 } ( \omega_1 ) ,
\\
Z^2\mathcal{M}_0 ( \Omega ; \omega_2 , \omega_1 )
= {} &
\beta \delta_{ \Omega , 0 }
\frac{ n }{ 2 }
G_0^{ - 1 } ( \omega_2 )
G_0^{ - 1 } ( \omega_1 ) ,
\\
Z^2\mathcal{C}_0 ( \Omega ; \omega_2 , \omega_1 )
= {} &
\beta \delta_{ \Omega , 0 }
\frac{ n }{ 2 } \left( 1 - n \right)
G_0^{ - 1 } ( \omega_2 )
G_0^{ - 1 } ( \omega_1 ) ,
\end{align}
\end{subequations}
and
$ \mathcal{S}_0 = 0 $,  where $G_0 ( \omega ) = Z /(i \omega + \mu_0 )$ is the
propagator of the Hubbard atom.
As both $ \mathcal{M}_0 $ and $ \mathcal{C}_0 $ are finite and singular,
any single-channel truncation of the X-FRG flow is bound to miss important physics of the projected Hilbert space.
Therefore we keep all three channels.
To detect possible ordering instabilities,
it is however sufficient to retain only the most singular contribution to the flow in  each channel \cite{Husemann2009}.
In this approximation,
the channel flow equations subject to the Hubbard atom initial conditions \eqref{eq:U0} can still be solved analytically.  
Furthermore,
we heuristically incorporate a part of the neglected diagram involving the three-body vertex via a Katanin substitution \cite{Katanin04}.
Details of the solution of the flow equations are presented in the companion paper \cite{Arnold2025}.
In the rest of this Letter,
we showcase the main features of the phase diagram and the spectral function of the square lattice $ t $ model that is generated by the X-FRG flow.
To that end,
we solved the flow equations for $ \Sigma_\Lambda ( K ) $ and $ \delta \mu_\Lambda $ numerically 
on a grid with 190 $ \bm{k} $-points in the irreducible wedge of the Brillouin zone
and 100 positive Matsubara frequencies.

\begin{figure}%
\includegraphics[width=1\linewidth]{fig2.png}%
\caption{Brillouin zone plots of the static part of 
 the magnetic channel 
$ M_{ \Lambda = 1 } ( \bm{q} , 0 ; 0 , 0 ) / t $.
(a) $ n = 0.2 $, $ T = 0.14 t $ in the paramagnetic state;
(b) $ n = 0.56 $, $ T = 0.13 t $ showing the stripe instability;
(c) $ n = 0.59 $, $ T = 0.16 t $, where stripe and ferromagnetic instabilities compete; and
(d) $ n = 0.85 $, $ T = 0.65 t $ showing the instability toward the Nagaoka ferromagnet.}
\label{fig:channels}
\end{figure}%
\textit{Magnetic instabilities---}Both superconducting and charge channels remain relatively featureless and  at most
$ \mathcal{O} ( t ) $ 
throughout the entire parameter space \cite{Arnold2025}.
In contrast,
the magnetic channel,
shown in Fig.~\ref{fig:channels} for a representative selection of $ n $ and $ T $, 
develops pronounced instabilities for $ n \gtrsim 0.48 $ at low temperatures:
For $ 0.48 \lesssim n \lesssim 0.59 $ 
the system favors a stripe phase with ordering vector $ \bm{q} = ( \pi , 0 ) $.
This is consistent with the best known variational wavefunctions for the Nagaoka state \cite{vonderLinden1991, Wurth1995},
which become unstable at a critical hole doping because the gap of a $ ( \pi , 0 ) $ spin wave vanishes.
At $ 0.6 \lesssim n \lesssim 1 $ on the other hand,
the dominant instability is toward the kinetic (Nagaoka) ferromagnet with $ \bm{q} = ( 0 , 0 ) $,
in agreement with Monte Carlo \cite{Becca2001} and density-matrix renormalization group results \cite{Liu2012, Blesio2019}.
Note that the 
magnetic instabilities shift to higher temperatures as the density is increased.
We attribute this to the increasing importance of the kinematic interaction due to the restricted Hilbert space with increasing $ n $.
Shortly after encountering the magnetic instabilities,
the exponentially growing fluctuations lead to a breakdown of the numerics.
Also, for $ n \gtrsim 0.99 $,
our truncation is apparently insufficient to deal with the strong kinematic interactions, 
preventing us from accessing the regime $ T \lesssim t $ and observing any kind of instability.

Because the magnetic ordering breaks the continuous $ O ( 3 ) $ spin-rotation invariance of the $ t $ model,
we expect that the correlation length diverges as 
$ \xi \sim e^{ 2 \pi \rho_s / T } $ 
for $ T \to 0 $ \cite{Shenker1980, Kopietz1989, Sandvik2010},
with a finite spin stiffness $ \rho_s (n) \sim t $ that is generated by the flow.
We therefore extract the correlation length from the static magnetic susceptibility 
$ \chi ( \bm{q} , \Omega = 0 ) $ \cite{Sandvik2010, Schafer2020, Arnold2025}
and fit its divergence immediately prior to the breakdown of the numerics to this form
to extract the $ \rho_s ( n ) $ shown in Fig.~\ref{fig:phases}.
Remarkably,
$ \rho_s ( n ) $ displays two separate peaks in the ferromagnetic regime.
The qualitative change may be related to the transition from partially to fully polarized ferromagnet at 
$ n_{ c , 3 } \approx 0.85 $~\cite{Chiappe1993, Becca2001, Coleman2002, Liu2012, Blesio2019}.

\begin{figure}%
\includegraphics[width=1\linewidth]{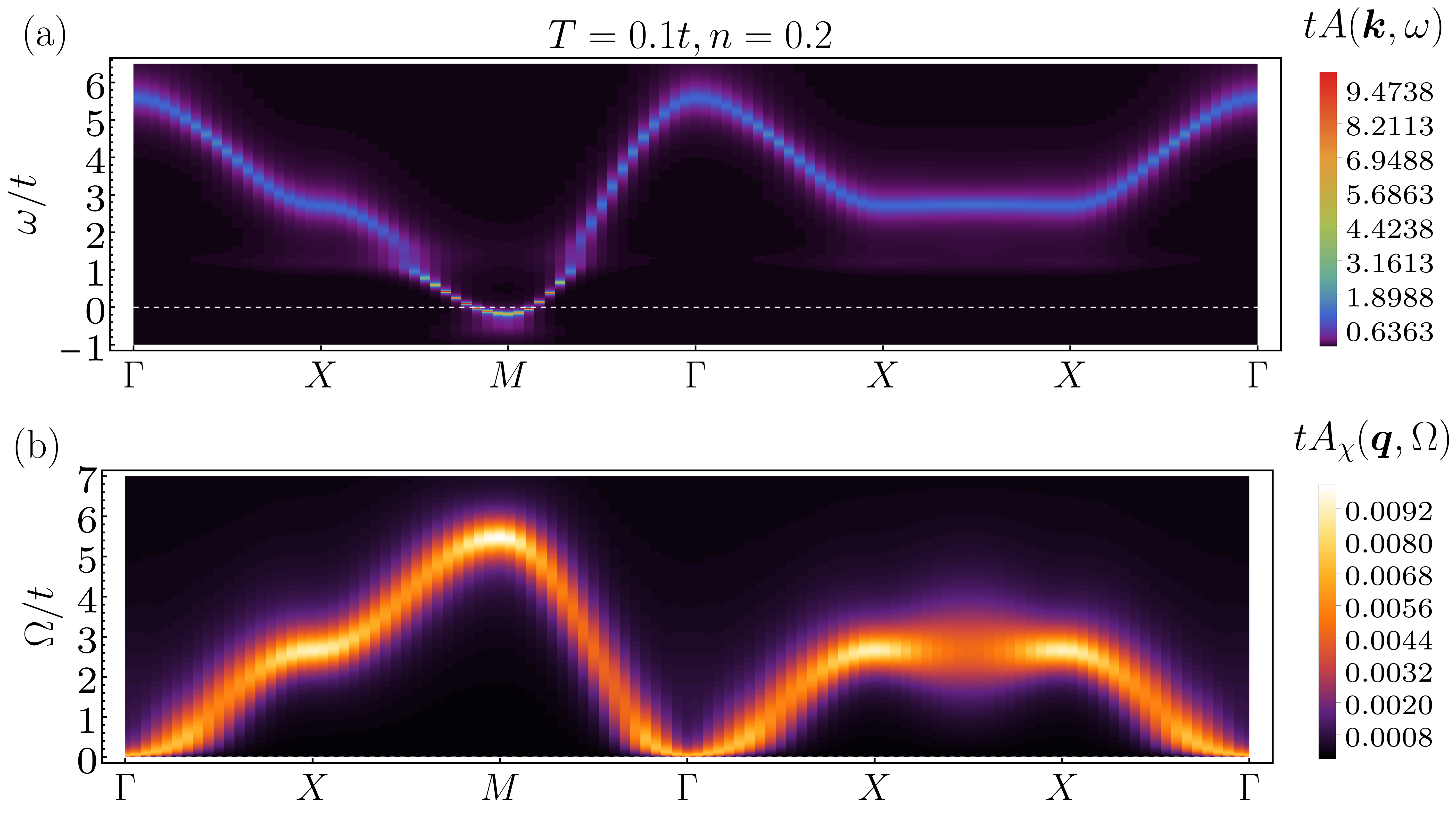}
\caption{Electronic (a) and magnetic (b) spectral functions 
along a high-symmetry path through the Brillouin zone, 
for 
$ n = 0.2 $ and 
$ T = 0.14 t $.
The dashed line in (a) marks the Fermi energy.
}
\label{fig:spectral_PM}
\end{figure}%
\textit{Spectral properties---}Since our X-FRG flow gives us access to the full holon self-energy $ \Sigma_\Lambda ( K ) $,
we can also address the electronic and magnetic spectral properties of the $ t $ model in the entire phase diagram.
To that end,
we analytically continue the holon propagator \eqref{eq:holon_propagator} and the dynamic magnetic susceptibility to real frequencies and compute the associated spectral functions,
given by
$ A ( \bm{k} , \omega )
= - \frac{ 1 }{ \pi }
\textrm{Im}\,  
G_{ \Lambda = 1 } ( \bm{k} , \omega ) |_{ i \omega \to \omega + i 0^+ } $ 
and
$ A_\chi ( \bm{q} , \Omega )
= \frac{ 1 }{ \pi }
\textrm{Im}\,  
\chi ( \bm{q} , \Omega ) |_{ i \Omega \to \Omega + i 0^+ } $,
respectively.
%
%
%
%
%
%
%
%
For the numerical analytical continuation,
we use Pad\'{e} approximants \cite{Beach2000, Schoett2016}.
In the paramagnetic phase at low densities shown in Fig.~\ref{fig:spectral_PM}(a),
we generally find a washed out band with sharp excitations around the Fermi edge,
indicating a Fermi liquid.
This is accompanied by a broad but well-defined paramagnon band in the spin response; see Fig.~\ref{fig:spectral_PM}(b).

\begin{figure}%
\includegraphics[width=1\linewidth]{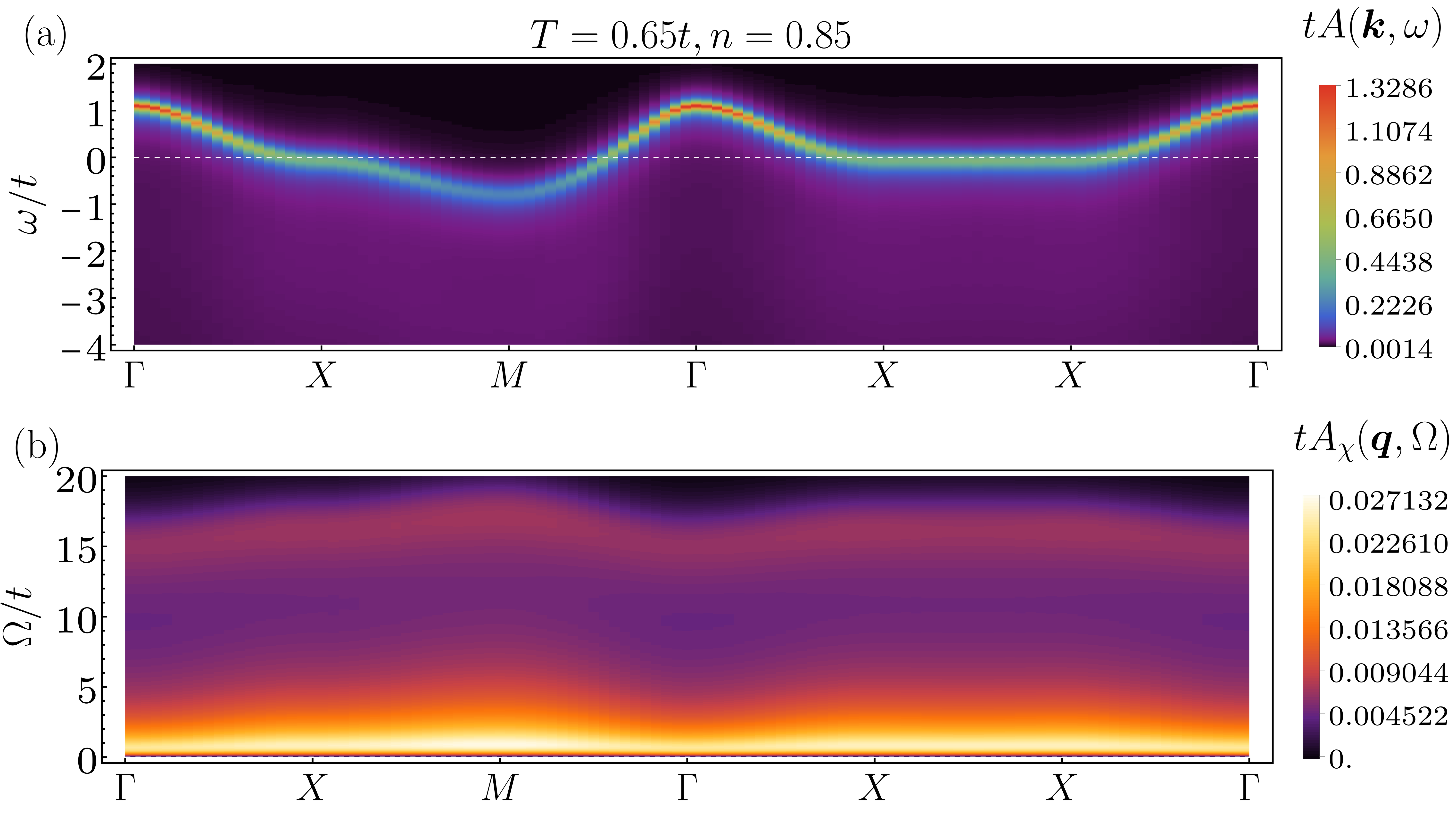}
\caption{Electronic (a) and magnetic (b) spectral functions 
along a high-symmetry path through the Brillouin zone, for 
 $ n = 0.85 $ and $ T = 0.65 t $
The dashed line in (a) marks the Fermi energy.
}
\label{fig:spectral_FM}
\end{figure}%
The most important physical results of our work are the   
spectral functions of the $t$ model in the Nagaoka regime ($ n > n_{c,2}$) shown in Fig.~\ref{fig:spectral_FM}.
In contrast to the low-density regime governed by single-particle excitations,
they exhibit several striking features indicative of strong-correlation cooperative physics.
Similar to other examples of correlation-induced ferromagnetism \cite{Kanamori1963, Linden1991, Mielke1991a, Mielke1991b, Tasaki2003, Hu2025}, 
the ferromagnetic instability coincides with the Fermi energy approaching the van Hove singularity of the narrow band.  
Let us point out two additional key features:
first,
upon approaching the Nagaoka state,
the electronic spectral function develops a almost flat but broadened band, 
accompanied by a significant shift of spectral weight to negative energies. 
The spectral broadening at the Fermi energy indicates the absence of well-defined quasi-particles 
and is one of the key features of an incoherent non-Fermi liquid
which has also been called a bad metal \cite{Park2008,Haule2003,Wang2018}.
The marked asymmetry between particle- and hole-like excitations as well as the weak momentum dependence of the spectrum are a consequence of the Nagaoka physics:
opposed to particles,
holes can lower their kinetic energy by binding a ferromagnetic bubble.
This gives rise to a continuum of Nagaoka polarons \cite{White2001, Maska2012, Lebrat2024, Prichard2024, Samajdar2024a} below the single-particle band that individually carry only little spectral weight \cite{Brinkman1970}.
As the polaron size is limited by the magnetic correlation length,
the hole dynamics at large $ n $ is long-range,
suppressing the $ \bm{k} $-dependence of the spectrum.
The magnon spectrum in Fig.~\ref{fig:spectral_FM}(b) likewise reflects the absence of well-defined single-particle excitations.
It exhibits only a weakly dispersive,
diffusive continuum that we associate with the multi-particle polaron states 
that also constitute the hole continuum in Fig.~\ref{fig:spectral_FM}(a).
\begin{figure}%
\includegraphics[width=1\linewidth]{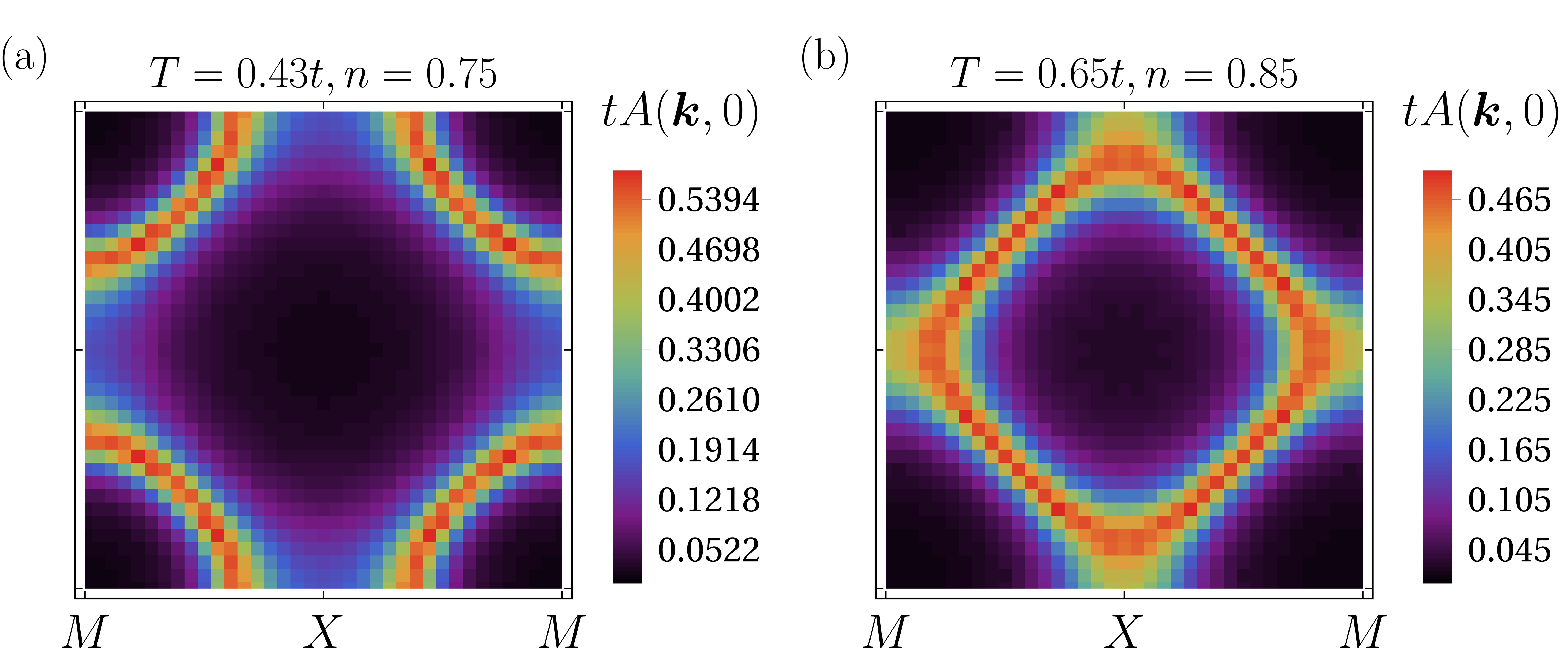}
\caption{Fermi surfaces determined via the spectral function $t A( \bm{k}, \omega = 0) $ for (a) $ n = 0.75 $ at $ T = 0.43 t $ and 
(b) $ n = 0.85 $ at $ T = 0.65 t $.}
\label{fig:Fermisurface}
\end{figure}%
The second striking feature is that 
the topology of the Fermi surface changes at some critical density 
$ n_{c,3} \approx 0.85 $ such that for 
$ n < n_{c,3} $ 
the Fermi surface is particle-like, 
while for 
$ n \ge n_{c,3} $ 
the Fermi surface is hole-like and consists of four disconnected arcs  in the first Brillouin zone;
see Fig.~\ref{fig:Fermisurface}.  
In other words, 
we find two distinct ferromagnetic regimes 
separated by a Lifshitz transition.
This accompanies the emergence of the second peak in $ \rho_s (n) $ shown in Fig.~\ref{fig:phases},
suggesting that the Lifshitz transition governs the transition from partially to fully polarized ferromagnet~\cite{Chiappe1993, Becca2001, Coleman2002, Liu2012, Blesio2019}.
\begin{figure}%
\includegraphics[width=1\linewidth]{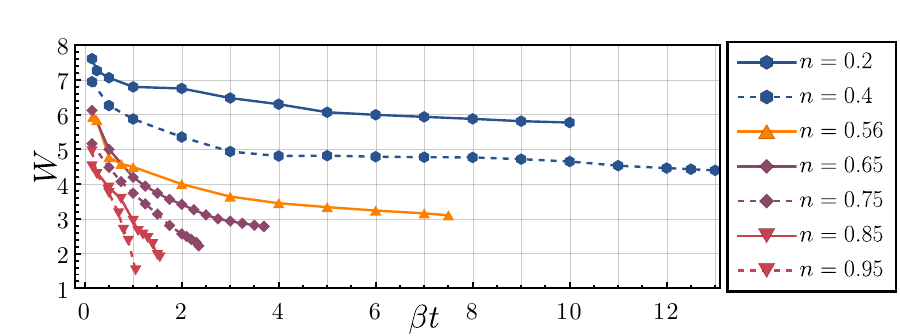}
\caption{Bandwidth $ W $ as function of inverse temperature $\beta t$ for various densities. 
The coloring corresponds to the extrapolated magnetic ground state: 
Blue: paramagnet; 
orange: stripe antiferromagnet; 
purple: FM1; 
red: FM2.}%
\label{fig:Bandwidth}%
\end{figure}%
Moreover,
from the $ T \to 0 $ extrapolation of the electronic bandwidth $ W $ shown in Fig.~\ref{fig:Bandwidth},
we conclude that the two different ferromagnetic regimes are connected to distinct ground states:
At densities $ n \lesssim 0.75 $,
$ W $ approaches a strongly renormalized but finite value.
Above the Lifshitz transition on the other hand,
$ W $ collapses drastically,
indicating that the band becomes truly flat, 
$ W \to 0 $, 
for $ T \to 0 $.
The appearance of an (almost) flat band is consistent with the fermion condensation scenario
proposed in 1990 by Khodel and Shaginyan \cite{Khodel1990} which has been  further developed in Refs.~[\onlinecite{Volovik1991, Shaginyan2010, Khodel2020}], in spite of
prominent criticism \cite{Nozieres1992} of the original mean-field 
calculation \cite{Khodel1990}. Our non-perturbative calculation  demonstrates that
extremely correlated fermions can indeed support flat bands.

\textit{Conclusions---}In this Letter we have shown that  the X-FRG
opens the door for the FRG  to reach the regime of extremely strong correlations
and therefore should be considered as a truly  transformative  methodological
advance.
We have used this method to calculate the phase diagram and the electronic and magnetic spectral functions
of the Hubbard model at $ U = \infty $,
which changes from a Fermi liquid governed by single-particle excitations at low densities to a cooperative magnet dominated by long-range multi-particle physics at high densities.
In particular, we predict the existence of an antiferromagnetic stripe phase at intermediate densities and have shown that in the Nagaoka regime the spectral function 
exhibits a flat band with anomalously large damping, in agreement with the bad metal phenomenology \cite{Park2008,Haule2003,Wang2018}. Moreover, we have detected a Lifshitz transition of the Fermi surface in the high-density (Nagaoka) regime 
that is accompanied by qualitative changes in the magnetic fluctuations and the electronic bandwidth.
It is straightforward to improve upon our truncation along the lines established for fermionic FRG \cite{Lichtenstein2017, Vilardi2017, Honerkamp2018, Vilardi2019, Ehrlich2020, Hille2020, Honerkamp2022}.
In future work,
the X-FRG approach could be extended to more general lattices and interactions,
as well as to more realistic models such as the $ t $-$ J $ model \cite{Rueckriegel2023}.
Moreover,
the lineshapes shown in Figs.~\ref{fig:spectral_PM} and \ref{fig:spectral_FM} can be useful to understand future experiments probing the spectral functions in extremely strongly correlated electronic systems realized in optical lattices \cite{Dehollain2019,Lebrat2024}.

\textit{Acknowledgments---}This work was financially supported by the Deutsche Forschungsgemeinschaft  (DFG, German Research Foundation) through Project No. 431190042. We thank G. E. Volovik for useful correspondence.

\textit{Data availability---}The data that support the findings of this study were generated and analyzed with Wolfram Mathematica. A custom Mathematica notebook containing all scripts is publicly available~\cite{gude}.

\end{document}